\documentstyle{l-aa}
\input{psfig.txt}

\begin{document}
%\double
\thesaurus{11.03.4;11.04.1}

\title{Optical spectroscopy of galaxies in the direction of the Virgo cluster.
\thanks{Based on observations obtained with the Loiano telescope belonging to the 
University of Bologna (Italy), with the OHP, operated by the French CNRS and with 
the G. Haro telescope of the INAOE (Mexico)}
}

\author{Giuseppe Gavazzi\inst{1}\and
Christian Bonfanti\inst{1}\and
Paola Pedotti\inst{1}\and
Alessandro Boselli\inst{2}\and
Luis Carrasco\inst{3,4}
}
\offprints{Gavazzi@uni.mi.astro.it}
\institute{Universit\`a degli Studi di Milano - Bicocca, P.zza dell'Ateneo Nuovo 1, 20126 M
ilano, Italy. 
\and
Laboratoire d'Astronomie Spatiale, Traverse du Siphon, F-13376 Marseille Cedex 12, France.
\and
Instituto Nacional de Astrof{\'i}sica, Optica y Electr\'onica,
Apartado Postal 51. C.P. 72000 Puebla, Pue., M\'exico
\and
Observatorio Astron\'omico Nacional/UNAM, Ensenada B.C., M\'exico
}
\maketitle
\markboth{Gavazzi et al.}{Virgo_cluster}

\begin{abstract}
Optical spectroscopy of 76 galaxies, 48 of which are projected in the direction
of the Virgo cluster and 28 onto the Coma--A1367 supercluster, is reported.
Adding these new measurements to those found in the literature, 
the redshift completeness in the Virgo region becomes 92 \%
at $\rm B_T\leq16.0$ and 68 \% at $\rm B_T\leq18.0$.
The one of CGCG galaxies in the direction of the  
Coma--A1367 supercluster becomes 98 \%.
The Virgo cluster membership estimates obtained on morphological
grounds by Binggeli et al. (1985) are confirmed in all cases.
However, several "possible members" classified as BCD (if in the
cluster) are found instead to be giant emission-line galaxies in the
background of the Virgo cluster

\end{abstract}

\keywords{galaxies: redshift; galaxies: large-scale structure; galaxies: clusters: individual: Virgo}

\section{Introduction}

The Virgo cluster, the nearest rich cluster of galaxies in the northern 
hemisphere, was surveyed with unprecedented resolution and completeness in
the extensive photographic survey carryed out with the DuPont telescope at
Las Campanas. Based on this material 
Binggeli et al. (1985) compiled  
the Virgo Cluster Catalogue (VCC), which lists 
2096 galaxies brighter than  $\rm B_T\le$20.0 (1323 of which constitute
a complete subsample to $\rm B_T\le$18.0). 
This work has served to many important developements, in particular for mapping
for the first time the
luminosity function of a cluster down to $M_p=$-13.1 (assuming a distance modulus of 
Virgo $\mu=-31.1$, corresponding to the distance of 17 Mpc; see Gavazzi et al. 1999a) 
and for comparing the various luminosity functions across the entire Hubble sequence 
(Sandage, Binggeli, \& Tammann, 1985).
In absence of a complete redshift information (by the time of its publication
only about 30\% VCC galaxies had an actual redshift measurement),
the cluster membership was assigned to the individual objects
on a purely morphological (surface brightness) basis.
Since then the number of available redshifts
either from optical (e.g. Binggeli et al. 1993, Grogin, Geller, \& Huchra 1998; hereafter GGH98) or from radio spectroscopy
(e.g. Haynes \& Giovanelli 1986, Hoffman et al. 1987, Hoffman et al. 1989,
Hoffman et al. 1995, Magri 1994) has increased considerably.
Nonetheless the present redshift completeness in this region is still only 65\% at $\rm B_T\le$18.0.

A better redshift completeness exists in the 
Coma--A1367 supercluster region $11^h30^m<\alpha<13^h30^m; 18^{\circ}<\delta<32^{\circ}$,
limited however to the shallower magnitude limit of 15.7 of the CGCG catalogue (Zwicky et al.
1961-68). 
Gavazzi et al. (1999b) counted 1068 redshift measurements out of 1127 CGCG galaxies
listed in this region.
With the aim of contributing with new redshift measurements in these regions
we undertook the spectroscopic survey presented in this paper, which was carryed
out during marginally photometric nights.
The observations and data reduction are presented in section 2.
The new redshifts are given and discussed in section 3. 

\begin{table*}
\caption{The spectrograph characteristics}
\label{Tab1}
\[
\begin{array}{p{0.15\linewidth}ccccccc}
\hline
\noalign{\smallskip}
{Telescope} & run & {Spectrograph} & {dispersion} & {coverage} & {CCD type} & {pix} \\ 
    &        &   &  {\rm \AA/mm}    &  {\rm \AA}     &           & {\mu m} \\
\noalign{\smallskip}
\hline
\noalign{\smallskip}
Loiano & Jan-Feb~1999 &  BFOSC  &  198 &  4060-7900  &  1024\times1024~TH    &  19 \\
Loiano & Jan-Feb~2000 &  BFOSC  &  198 &  3600-8900  &  1340\times1300~EEV    &  20 \\
Cananea & Apr~2000  & LFOSC &  228 &  4000-7100  &  576\times384~TH      &  23 \\ 
OHP   & Mar~1999  & CARELEC & 133 &  3200-7100  &  2048\times1024~EEV   &  13.5 \\ 
\noalign{\smallskip}
\hline
\end{array}
\]
\end{table*}

\section{Observations and data reduction}

Galaxies in the present study were primarily selected among
the objects brighter than $\rm B_T\leq17.0$ in the
VCC Catalogue of Virgo Cluster galaxies by Binggeli et al. (1985).
CGCG (Zwicky et al. 1961-68) galaxies in the region
$11^h30^m<\alpha<13^h30^m; 18^{\circ}<\delta<32^{\circ}$, containing
the Coma--A1367 supercluster, were also selected as filler objects.
Long-slit, low dispersion spectra of 76 galaxies were obtained in several 
observing runs since 1999 using the imaging spectrographs BFOSC and LFOSC
attached to the Cassini 1.5 m telescope at Loiano (Italy), to
the 2.1 m telescope of the Guillermo Haro Observatory at Cananea (Mexico), respectively,
and with the CARELEC spectrograph (Lemaitre et al. 1990) attached to the 1.92 m telescope of 
the Observatoire de Haute Provence (OHP) (France).

Table 1 lists the characteristics of the instrumentation
in the adopted set-up.

The observations at Loiano were performed using a 2.0 or 2.5 arcsec slit, 
depending on the seeing conditions, generally oriented E-W.
Every galaxy spectrum was preceded and followed by an exposure of
a HeAr lamp to secure the wavelength calibration.
The exposure time ranged between 20 and 90 min (1999 run) according to 
the brightness of the target objects, or 15 min (2000 run) owed
to the much higher quantum efficiency of the new EEV detector. 

The observations at Cananea were carried out
with a 1.9 arcsec slit, generally oriented N-S. 
Every galaxy spectrum was preceded and followed by an exposure of
a XeNe lamp to secure the wavelength calibration.
The exposure time ranged between 20 and 40 min according to 
the brightness of the target objects.

The observations at OHP were carried out
with a 2.5 arcsec fixed slit, generally oriented E-W. 
Every galaxy spectrum was preceded and followed by an exposure of
a HeAr lamp to secure the wavelength calibration.
The exposure time ranged between 20 and 30 min according to 
the brightness of the target objects.
In all runs the observations were obtained in nearly photometric conditions,
with thin cirrus. The
orientation of the slit was modified from the set-up given above when two 
adiacent objects were observable in the same exposure.

The data reduction was performed in the IRAF-PROS environment.
After bias subtraction, when 3 or more frames of the same target
were obtained, these were combined (after spatial alignment)
using a median filter to help cosmic rays removal. 
Otherwise the cosmic rays were removed under visual inspection.
The wavelength calibration was checked on known sky lines. These were
found within $\sim$ 1 $\rm \AA$ from their nominal value, providing an
estimate of the systematic uncertainty on the derived velocities of $\sim$ 50 $\rm km~ s^{-1}$. 
After subtraction of the sky background, one-dimensional spectra were
extracted from the frames.
These spectra were analyzed with either of two methods:
 
1) {\bf individual line measurement}: all spectra taken at Loiano 2000
were inspected and emission/absorption
lines were identified. Emission lines include H$\alpha$, [NII] and [SII].
Absorption lines include the MgI, Ca-Fe and Na.
The galaxy redshift was obtained from these individual measurements.
If more than one line was identified, the galaxy redshift was derived as
the weighted mean of the individual measurements, with weights 
proportional to the line intensities.
 
2) {\bf cross correlation technique}: spectra obtained in all the remaining runs 
were analyzed using the cross-correlation technique of 
Tonry \& Davis (1979). 
This method is based on a "comparison" between the spectrum of a galaxy whose
redshift is  to be determined, and a fiducial spectral template 
of a galaxy (or star) of appropriate spectral type to contain the wanted 
absorption/emission lines.
The basic assumption behind this method is that the spectrum of a
galaxy is well approximated by the spectrum of its stars, modified
by the effects of the stellar motions inside the galaxy
and by the systemic redshift.
For this purpose high signal-to-noise spectra 
were taken of four template galaxies: M105 and M32 (absorption lines) 
and VCC1554 and IC342 (emission lines), which were converted to the restframe $\lambda$.
The observed redshifts ($V_{obs}$) were not transformed to Heliocentric.

\section{Results}
\begin{figure*}
\centerline{
\psfig{figure=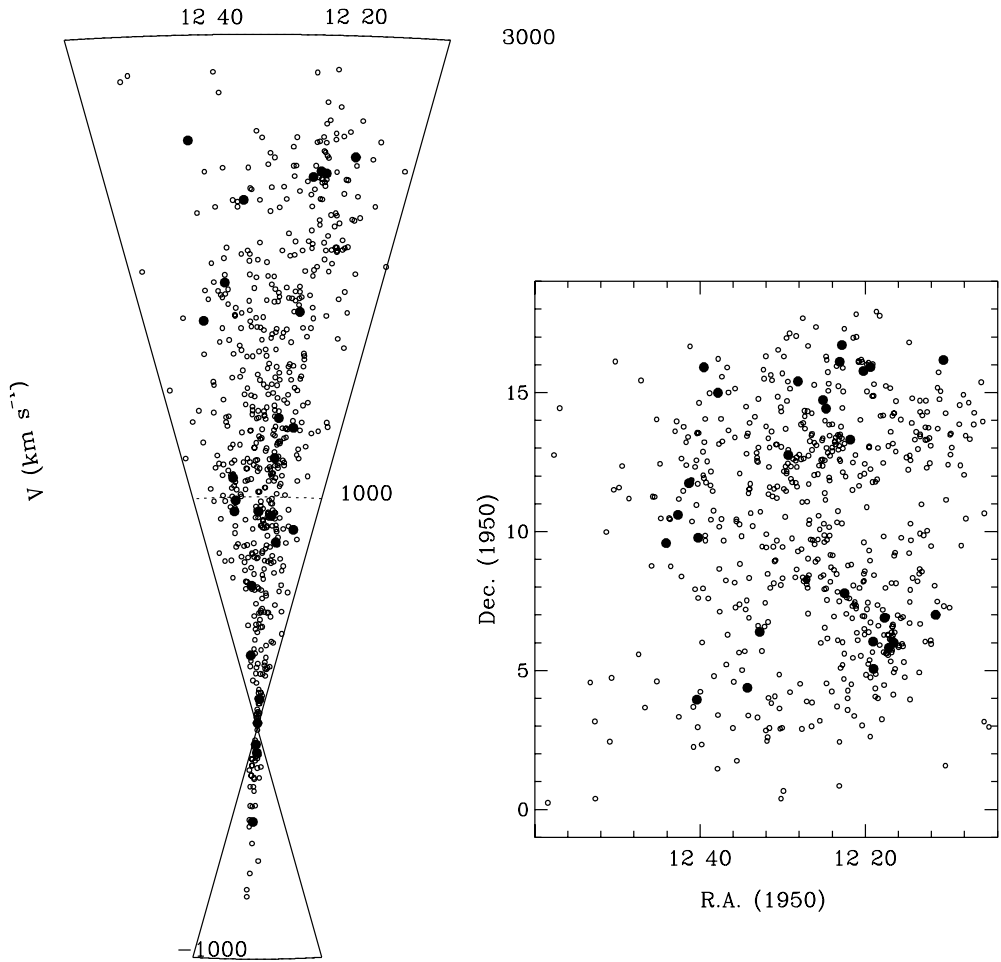,width=14truecm,height=14truecm}}
\caption{The distribution in celestial coordinates of 639 Virgo galaxies
with $V\leq 3000$~$\rm km~ s^{-1}$
(right) and a wedge diagram (left). The filled symbols represent measurements obtained in the
present work.}
\end{figure*}
The velocity measurements obtained in this work are listed
in Table 2 (Virgo) and 3 (Coma) as follow: \newline
Column 1: the CGCG  (Zwicky et al. 1961-68) or VCC (Binggeli et al. 1985) designation.\newline
Columns 2, 3: (B1950) celestial coordinates, measured with few
arcsec uncertainty.\newline
Column 4: morphological type as given in the VCC.\newline
Column 5: photographic magnitude.\newline
Column 6, 7: observed recessional velocity with uncertainty derived in this work.
The latter quantity includes only statistical errors. The global uncertainty
can be derived by adding in quadrature the systematic error of 50 $\rm km~ s^{-1}$
due to the uncertainty in the absolute wavelength calibration. 
Several absorption line spectra have statistical errors up to 900 $\rm km~ s^{-1}$,
reflecting a lack of strong features. These redshifts, however, are sufficient to derive a membership.
\newline
Column 8: type of lines (A=absorption; E=emission).\newline
Column 9: observing run (L99=Loiano 1999, L00=Loiano 2000, Can=Cananea 2000, 
OHP=OHP 1999). \newline
Column 10: old membership as given in the VCC (bk=background, m=member, -=possible member)(only for Table 2). \newline
Column 10a: new membership. \newline
Column 11, 12: previously available redshift, with reference. 

\begin{table*}
\caption{Parameters of the observed Virgo galaxies}
\label{Tab3}
\[
\begin{array}{p{0.15\linewidth}ccccccccccccc}
\hline
\noalign{\smallskip}
Gal.  & RA(1950) & Dec(1950)    & Type  & \rm B_T  & V &  \pm & Lines &  Run & Memb & new Memb & v_{alt} & ref\\ 
      &~h~m~s    &~^o~'~" & &  \rm mag & \rm km~s^{-1}    &   \\
 (1)    & (2) & (3) & (4) & (5) & (6) & (7) & (8) & (9) & (10) & (10a) & (11) & (12)\\
\noalign{\smallskip}
\hline
\noalign{\smallskip}
VCC0007  & 120645.60 & 114230.0  &  Sc   & 15.04 & 18675 & 425 & A & Can  & bk &  bk  &     &       \\
VCC0014  & 120717.80 & 113205.0  &  BCD? & 16.50 & 17891 &  35 & E & L00  & -  &  bk  &     &       \\
VCC0019  & 120740.80 & 132800.0  &  BCD? & 16.50 &  6803 &  99 & E & L00  & -  &  bk  &     &       \\
VCC0045  & 120934.70 & 152315.0  &  BCD? & 16.00 & 15236 &  36 & E & L00  & -  &  bk  &     &       \\
VCC0064  & 121008.50 & 114955.0  &  Sab  & 15.04 & 18142 & 336 & A & Can  & bk &  bk  &     &       \\
VCC0074  & 121031.80 & 161024.0  &  BCD? & 16.30 &   861 &  78 & E & Can  & -  &  m   &     &       \\
VCC0099  & 121128.90 &  70004.0  &  Sa?  & 14.81 &  2476 & 214 & E & L99  & -  &  m   & 2444 & GGH98  \\
VCC0196  & 121400.00 &  94624.0  &  BCD? & 16.50 & 13024 &  57 & E & L00  & -  &  bk  &     &       \\
VCC0225  & 121439.00 &  83612.0  &  BCD? & 17.00 & 21345 & 126 & E & L00  & -  &  bk  &     &       \\
VCC0249  & 121509.00 & 133930.0  &  Sa   & 14.61 &  7491 &  61 & E & Can  & bk &  bk  &     &       \\
VCC0323  & 121633.20 &  60012.0  &  Sa   & 14.91 &  2402 & 358 & A & L99  & -  &  m   & 2756 & GGH98  \\
VCC0362  & 121709.00 &  54856.0  &  Sa   & 14.51 &  1300 & 304 & A & L99  & -  &  m   & 1536 & GGH98  \\
VCC0397  & 121739.00 &  65402.0  &  dE?  & 15.00 &  2411 & 809 & A & L99  & -  &  m   & 2495 & GGH98  \\
VCC0482  & 121900.80 &  50324.0  &  S0a  & 14.77 &  1802 & 709 & A & L99  & -  &  m   & 2170 & GGH98  \\
VCC0486  & 121903.80 &  60235.0  &  S0a  & 14.50 &  2386 & 252 & A & L99  & -  &  m   & 2498 & GGH98  \\
VCC0510  & 121922.80 & 155518.0  &  dE   & 15.13 &   804 & 151 & A & Can  & m  &  m   &     &       \\
VCC0541  & 121945.00 &  43348.0  &  BCD  & 16.00 & 23511 &  50 & E & L00  & -  &  bk  &     &       \\ %no
VCC0573  & 122009.60 &  55454.0  &  Sc   & 15.20 & 23083 & 189 & E & Can  & bk &  bk  & 23083 & NED \\
VCC0583  & 122014.40 & 154636.0  &  Im   & 15.76 &   -72 & 475 & A & L99  & m  &  m   &     &       \\
VCC0723  & 122149.80 & 131824.0  &  dS0? & 15.04 &   125 &  50 & A & L00  & -  &  m   &     &       \\
VCC0762  & 122230.00 &  74660.0  &  dE   & 15.30 &  1341 & 211 & A & Ohp  & m  &  m   &     &       \\
VCC0794  & 122250.40 & 164224.0  &  dS0  & 15.50 &   918 & 817 & A & Ohp  & m  &  m   &     &       \\
VCC0817  & 122306.00 & 160642.0  &  dE   & 15.00 &  1168 & 153 & A & Can  & m  &  m   &     &       \\
VCC0991  & 122445.90 & 142525.0  &  dE   & 14.70 &  -406 & 239 & A & L99  & m  &  m   &     &       \\
VCC1028  & 122506.60 & 144360.0  &  dS0? & 15.70 &    21 & 158 & A & Ohp  & -  &  m   &     &       \\
VCC1174  & 122645.80 & 101246.0  &  BCD? & 15.50 & 11840 &  52 & E & L99  & -  &  bk  &     &       \\
VCC1270  & 122743.80 &  84800.0  &  Sa   & 15.00 & 11687 & 440 & A & Can  & bk &  bk  &     &       \\
VCC1304  & 122809.00 & 152412.0  &  dS0  & 15.50 &  -108 & 294 & A & L99  & m  &  m   &     &       \\
VCC1389  & 122919.80 & 124530.0  &  dE   & 15.91 &   936 & 193 & A & Can  & m  &  m   &     &       \\
VCC1395  & 122923.40 &  85248.0  &  dE?  & 16.20 & 22900 & 100 & E & L00  & -  &  bk  &     &       \\
VCC1423  & 122942.60 &  31630.0  &  BCD? & 16.00 & 13079 &  98 & A & Can  & -  &  bk  &     &       \\
VCC1608  & 123247.60 &  62325.0  &   E   & 14.20 &  2285 & 193 & A & L99  & -  &  m   & 2464 &  GGH98 \\
VCC1643  & 123321.00 &  60212.0  &  S0   & 15.20 & 12509 & 258 & A & Ohp  & -  &  bk  & 12563 & GGH98 \\
VCC1671  & 123359.60 &  62641.0  &  dS0  & 14.80 & 11608 & 809 & A & L99  & -  &  bk  & 11846 & GGH98 \\
VCC1687  & 123416.20 &  42242.0  &  dE   & 15.10 &   616 & 122 & A & Can  & -  &  m   &     &       \\
VCC1836  & 123749.80 & 145930.0  &  dS0  & 14.54 &  1927 & 148 & A & Can  & m  &  m   &     &       \\
VCC1849  & 123803.60 &  94942.0  &  BCD? & 16.20 & 15905 &  50 & E & L00  & -  &  bk  &     &       \\
VCC1906  & 123932.20 & 155438.0  &  S0   & 15.70 &   314 & 138 & A & Can  & -  &  m   &     &       \\
VCC1927  & 124005.40 & 105024.0  &  Sc   & 14.91 & 20085 & 180 & A & Can  & bk &  bk  &     &       \\
VCC1936  & 124014.40 &  94654.0  &  dS0  & 15.68 &   985 & 276 & A & Ohp  & m  &  m   &     &       \\
VCC1947  & 124023.30 &  35701.0  &  dE   & 14.56 &  1083 & 405 & A & L99  & -  &  m   & 944 &  GGH98  \\
VCC1956  & 124036.00 &  35118.0  &  S..  & 15.10 & 14691 &  51 & E & L99  & -  &  bk  & 14659 & GGH98 \\
VCC1982  & 124119.20 & 114412.0  &  dE   & 15.30 &   938 & 464 & A & Ohp  & m  &  m   &     &       \\
VCC1997  & 124151.60 & 102742.0  &  Sb   & 15.10 &  9210 &  46 & E & Can  & bk &  bk  &     &       \\
VCC2015  & 124240.20 & 103554.0  &  BCD? & 16.20 &  2545 & 115 & E & L00  & -  &  m   &     &       \\
VCC2042  & 124407.20 &  93448.0  &  dE   & 14.84 &  1765 & 154 & A & Can  & m  &  m   &     &       \\
VCC2077  & 124604.50 & 110851.0  &  Sab  & 15.20 & 11860 & 225 & A & Can  & bk &  bk  &     &       \\
VCC2082  & 124727.60 & 113206.0  &  S..  & 15.30 &  7421 &  26 & E & Can  & bk &  bk  &     &       \\
\noalign{\smallskip}
\hline
\end{array}
\]
\end{table*}

\begin{table*}
\caption{Parameters of the observed Coma galaxies}
\label{Tab3b}
\[
\begin{array}{p{0.15\linewidth}cccccccccccc}
\hline
\noalign{\smallskip}
Gal.  & RA(1950) & Dec(1950)    & Type  & \rm B_T  & V &  \pm & Lines &  Run & Memb & v_{alt} & ref\\ 
      &~h~m~s    &~^o~'~" & &  \rm mag & \rm km~s^{-1}    &   \\
 (1)    & (2) & (3) & (4) & (5) & (6) & (7) & (8) & (9) & (10a) & (11) & (12) \\
\noalign{\smallskip}
\hline
\noalign{\smallskip}
 127-028  & 113851.25 & 250457.3  &  S0  & 15.60 &  3518 & 330 & A & L99  & fg   &     &       \\
 127-029N & 113859.62 & 260956.6  &   E  & 16.30 &  7407 &  14 & E & Can  & m    &     &       \\
 127-029S & 113900.06 & 260926.5  &   E  & 16.30 &  6927 & 113 & A & Can  & m    &     &       \\
  97-153W & 114513.24 & 184955.0  &  S.. & 16.30 & 11156 &  20 & E & Can  & bk   &     &       \\
  97-153E & 114515.08 & 184936.7  &  S.. & 16.30 & 20261 & 112 & E & Can  & bk   &     &       \\
 127-057S & 114550.62 & 260223.9  &  S.. & 16.50 & 13666 &  75 & E & Can  & bk   &     &       \\
 127-057N & 114551.69 & 260251.3  &  S.. & 16.50 & 13718 &  15 & E & Can  & bk   &     &       \\
 127-102  & 115130.07 & 232813.1  &   E  & 15.70 &  7799 & 139 & A & Ohp  & m    &     &       \\
 128-028W & 120405.12 & 260142.4  &   E  & 16.40 &  7353 & 198 & A & Can  & m    &     &       \\
 128-028E & 120406.94 & 260150.2  &   E  & 16.40 & 13788 & 143 & A & Can  & bk   &     &       \\
  98-088  & 120903.25 & 201024.8  &  S0  & 15.70 &  6564 & 215 & A & Ohp  & m    &     &       \\
 128-055  & 121122.87 & 220200.8  &  S0  & 15.70 &  7227 & 330 & A & L99  & m    &     &       \\
  98-120  & 121341.37 & 194405.9  &   E  & 15.70 & 13208 & 202 & A & Ohp  & bk   &     &       \\
  98-127  & 121409.85 & 183918.2  &   E  & 15.70 &  8954 & 203 & A & Ohp  & bk   &     &       \\
  99-013  & 121638.07 & 193306.1  &  Sc  & 15.70 &  7297 &  16 & E & Ohp  & m    &     &       \\
 128-083  & 121936.55 & 240846.0  &   E  & 15.70 & 10182 & 129 & E & Ohp  & bk   &     &       \\
 128-083E & 121939.64 & 240847.5  &   E  & 17.00 & 10292 & 164 & A & Ohp  & bk   &     &       \\
 128-085  & 122150.98 & 212612.2  &  Sc  & 15.60 &   914 &  21 & E & Ohp  & fg   &     &       \\
  99-066  & 122620.69 & 194525.9  &  Sb  & 15.70 & 13582 &  86 & E & Ohp  & bk   &     &       \\
 129-003  & 122631.44 & 245429.9  &  Sc  & 15.70 & 14572 &   9 & E & Ohp  & bk   &     &       \\
  99-067  & 122635.25 & 191652.4  &   E  & 15.70 & 14393 & 250 & A & Ohp  & bk   &     &       \\
 159-087E & 124810.87 & 274149.8  &  Sbc & 15.70 & 12286 &  16 & E & Ohp  & bk   &     &       \\
 160-036S & 125435.00 & 305820.9  &  S0  & 16.00 & 15302 & 177 & A & Can  & bk   &     &       \\
 160-036E & 125438.62 & 305831.5  &   E  & 16.50 & 14880 & 185 & A & Can  & bk   &     &       \\
 160-163S & 131035.69 & 272401.1  &   E  & 16.50 & 17929 & 178 & A & Can  & bk   &     &       \\
 160-163  & 131036.56 & 272421.7  &  S0a & 15.70 & 18015 &  72 & E & Can  & bk   &     &       \\
 161-029  & 131912.70 & 263359.0  &  Sb  & 15.70 &  4930 &  11 & E & Ohp  & m    &     &       \\
 161-061S & 132554.00 & 285542.1  &  S.. & 15.60 & 11281 &  32 & E & Can  & bk   & 11247 & G99 \\
 161-061N & 132554.69 & 285657.4  &   E  & 16.50 & 10564 & 165 & A & Can  & bk   &     &       \\
\hline
\end{array}
\]
\end{table*}

Fig. 1 gives a representation in celestial coordinates of 639 galaxies
in the VCC region with redshift  $V\leq 3000$~$\rm km~ s^{-1}$
(panel a) and a wedge diagram (in the same velocity window) is given in panel b.
Small symbols mark galaxies taken from the literature, filled circles 
mark the measurements obtained in this work.
Fig. 2 gives a representation in celestial coordinates of 913 galaxies
in the VCC region with redshift  $\leq 24000$~$\rm km~ s^{-1}$
(panel a) and a wedge diagram is given in panel b.
\begin{figure*}
\centerline{
\psfig{figure=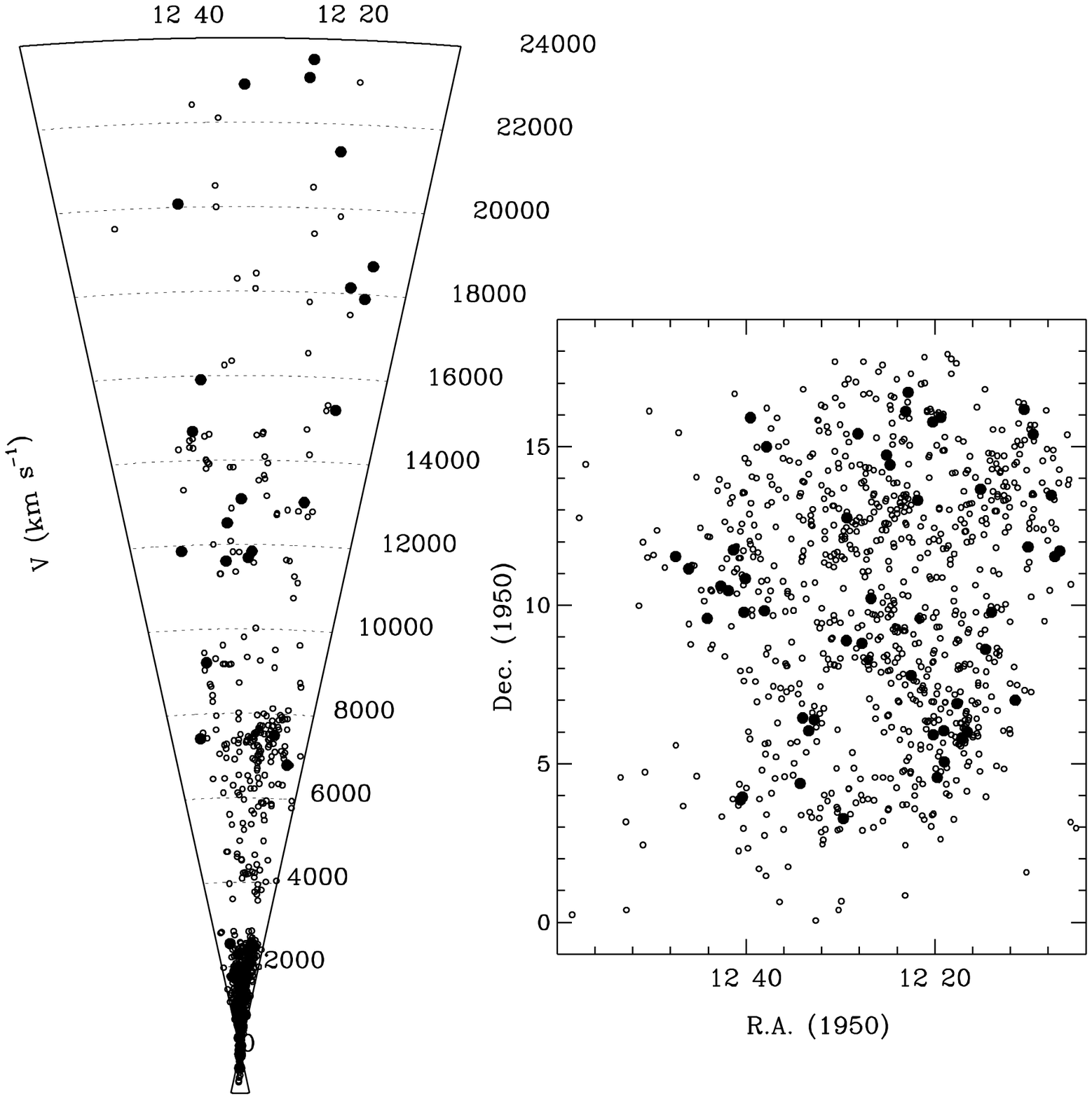,width=14truecm,height=14truecm}}
\caption{The distribution in celestial coordinates of 913 Virgo galaxies
with $V\leq 24000$~$\rm km~ s^{-1}$
(right) and a wedge diagram (left) (same symbols as in Fig. 1).}
\end{figure*}
Fig. 3 gives a representation in celestial coordinates of 1109 galaxies
in the Coma region with measured redshift 
(top) and a wedge diagram is given (bottom).
(Same use of symbols as in Figs. 1 and 2).
\begin{figure*}
\centerline{
\psfig{figure=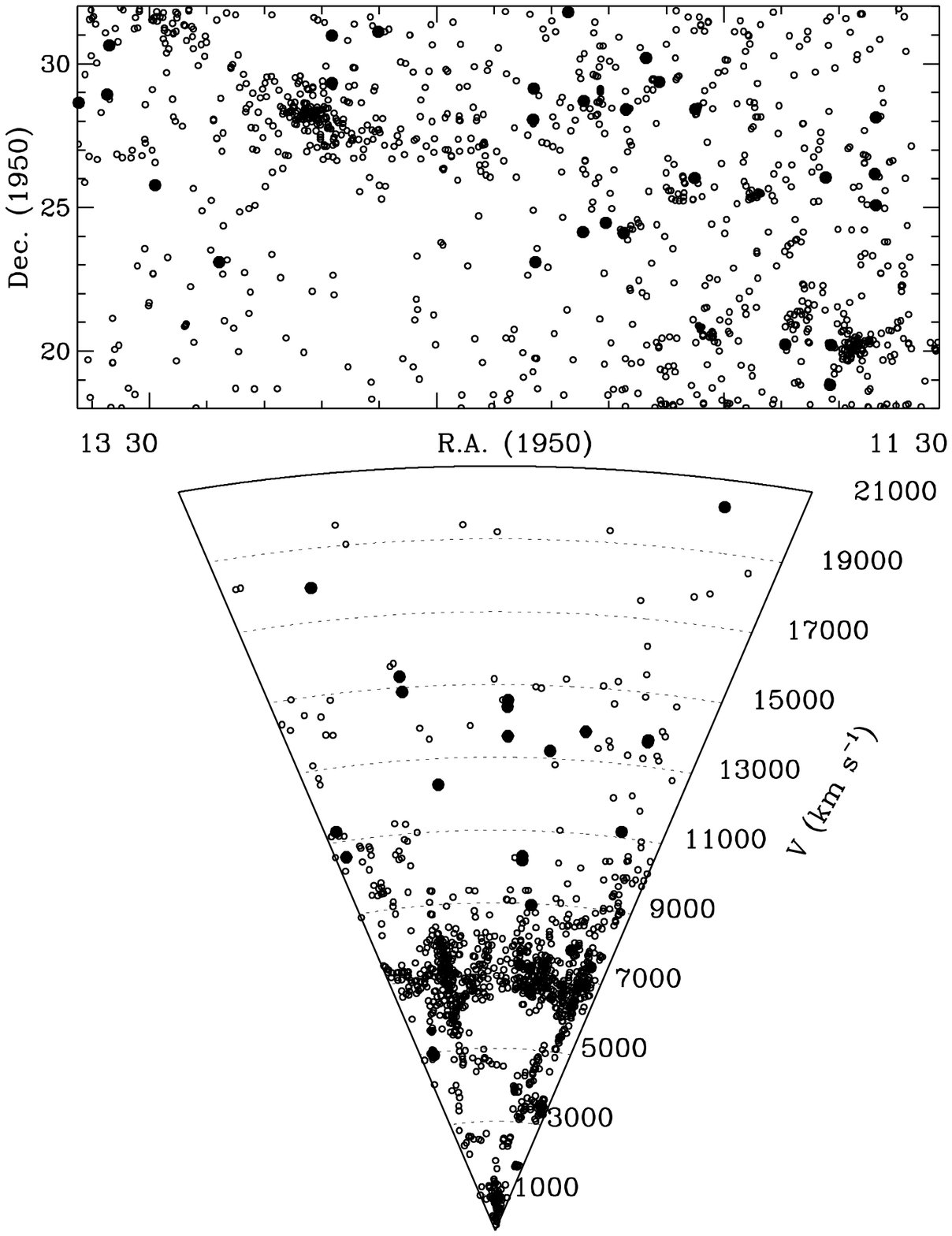,width=14truecm,height=14truecm}}
\caption{The distribution in celestial coordinates of 1109 galaxies 
in the direction of the Coma supercluster with measured redshift
(top) and a wedge diagram (bottom)(same symbols as in Fig. 1).}
\end{figure*}

\subsection{redshift completeness}

The VCC catalogue contains 2096 galaxies brighter than $\rm B_T=20.0$. Of these
only 913 have a redshift measurement so far. Even at brighter levels the
redshift information is far from complete (see Tab. 4 for details).
For example among the 849 galaxies with $\rm B_T<16.0$ there are still 69 with
no spectra available. It is not surprising, though, that such relatively 
bright objects remain unmeasured because, given the vicinity of the Virgo cluster,
galaxies in this magnitude range have low luminosities ($M_p>-15.0$). 
Given the known inverse proportionality between the luminosity and 
the surface brightness, they all correspond with extremely low surface brightness
galaxies, which are much more difficult to observe spectroscopically than
objects of similar magnitude which are further away. Furthermore, galaxies with missing
spectra are almost entirely dEs, thus with featureless spectra.
During the spectroscopical runs described in the present paper we have
tried, unsuccesfully, to measure several of these objects:
VCC 236 (dE), 452 (dE), 816 (dE), 1417 (dE), 1497 (dE), 1503 (dE), 1571 (dE), 1649 (dE), 
1755 (dE), 1825 (dIm/dE), 1945 (dE), 1991 (dE), 2083 (dS0) with integration
time $\sim$ 30 min. We point out to those observers who wish to obtain
succesfull spectra at 2m class telescopes to try with much longer exposures.
 
The redshift completeness is far better in the Coma-region, limited however to a brighter
magnitude.
Of the 1127 CGCG galaxies listed in the Coma region, 1082 have
$\rm B_T\le$15.7. Another 45 belong to multiple systems which
were split in their individual components, each
of them fainter than the catalogue limiting magnitude 15.7.
Only 2/1082 galaxies with $\rm B_T\le$15.7 and 18/1127 with $\rm B_T\le$16.5 
remain with unknown redshift, thus the sample is 98\% complete.

An interesting example of a strong emission-line object in the background of the Coma
supercluster is CGCG 127-057N which
was observed at Cananea. The spectrum of this galaxy (see Fig.4) 
shows strong Balmer and [OIII] lines and weak [NII] and [SII].
The corresponding metallicity derived from $[OIII]/H\beta$, as prescribed by
Edmunds \& Pagel (1984) is: 12+log(O/H)=8.16.
\begin{table}
\caption{redshift completeness in the Virgo cluster}
\label{Tab3}
\[
\begin{array}{p{0.15\linewidth}cccc}
\hline
\noalign{\smallskip}
{mag} & tot & with~z & \%  \\ 
\noalign{\smallskip}
\hline
\noalign{\smallskip}
$\leq$15.0 & 549 & 546 & 99\\
$\leq$16.0 & 849 & 780 & 92\\
$\leq$17.0 & 1064 &  864 &  81\\
$\leq$18.0 & 1323 & 903 & 68\\
$\leq$19.0 & 1704 & 913 & 53\\
$\leq$20.0 & 2096 & 916 & 43\\
\noalign{\smallskip}
\hline
\end{array}
\]
\end{table}

\subsection{membership}

\begin{figure}
\psfig{figure=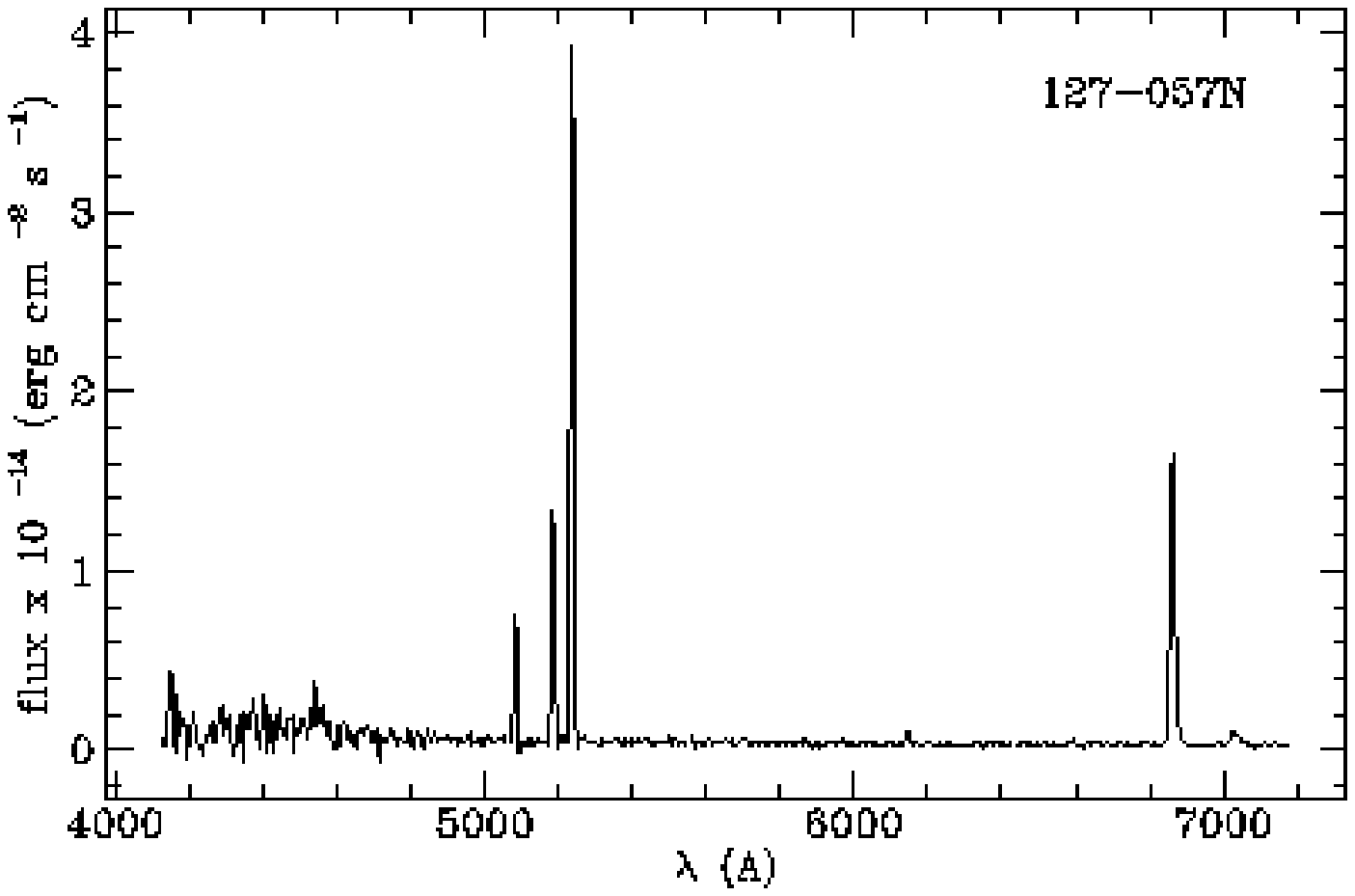,width=9truecm,height=7truecm}
\caption{The emission-line spectrum of 127-057N}
\end{figure}
The membership to the Virgo cluster given in the VCC (members, possible-members,
background) was estimatied on purely
morphological grounds (mostly on the surface brightness) by Binggeli et al. (1985).
The new redshifts presented in this paper, in conjunction with two other
recent sets of Virgo velocity measurements (Magri 1994) and 
GGH98), can be used to reassess this issue.
We find that all objects listed in the VCC as "members" are confirmed as such
($V<3000$~$\rm km~ s^{-1}$), stressing the high success-rate of the morphological
estimate.
"Possible members" are found with $V<3000$~$\rm km~ s^{-1}$ in 67 \% of cases, 
the remaining being background objects.
We have noticed that out of the 13 "Possible members" which are in fact
high redshift objects, 10 were classified as possible BCDs (BCD?). These appear to be
systematically emission-line
blue giant objects, not dwarfs, nor compact galaxies (see Tab. 2).

As an example we give in Fig. 5 the rotation curve of one of them: VCC1849,
which has a total rotational velocity up to 300 ~$\rm km~ s^{-1}$,
not typical of a dwarf galaxy. The rotation curve was derived at intermediate
dispersion during the Loiano 2000 run with the slit oriented along the galaxy major axis.
\begin{figure}
\psfig{figure=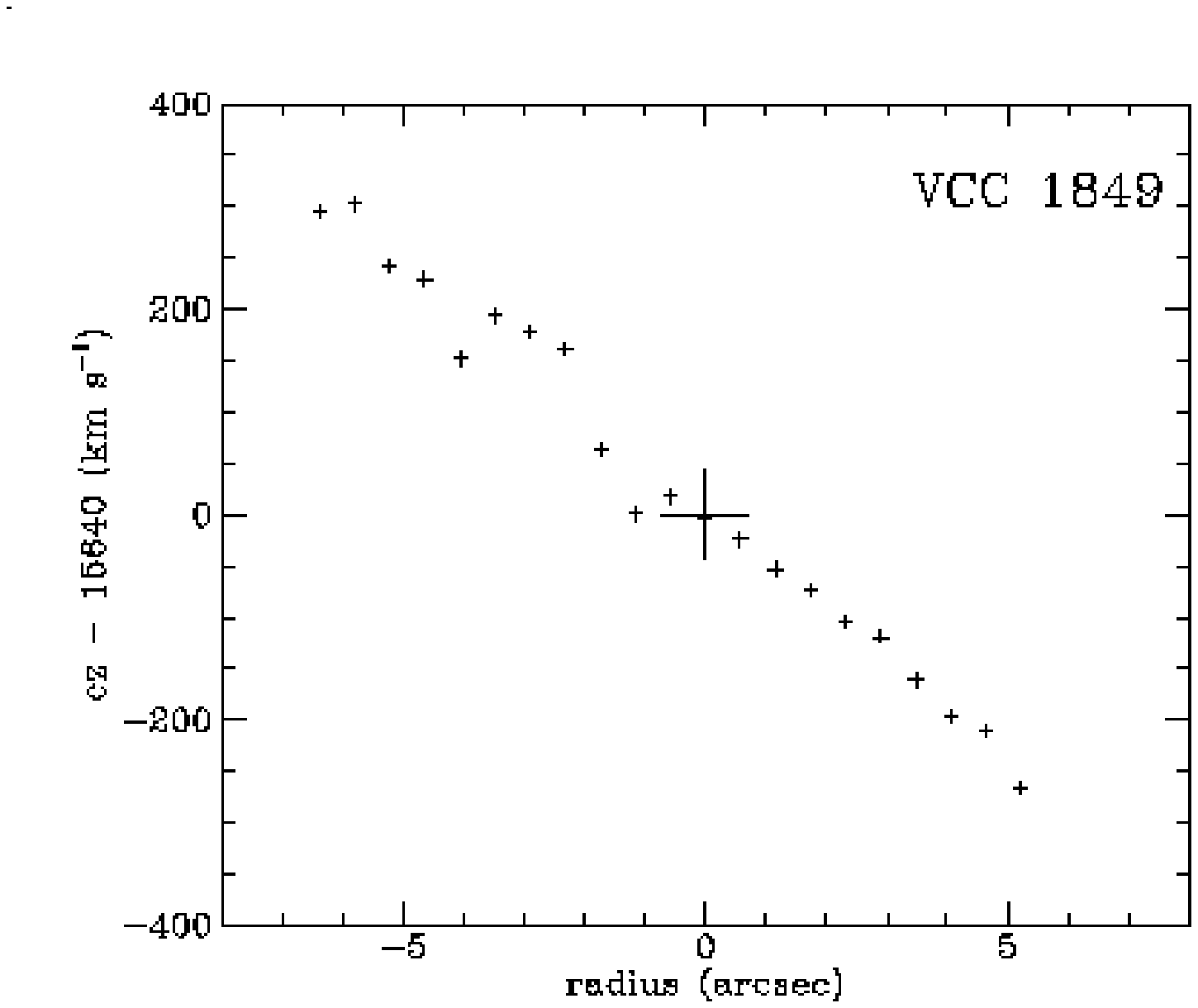,width=9truecm,height=7truecm}
\caption{The $H\alpha$ rotation curve of VCC1849}
\end{figure}
All objects classified as "Background" in the VCC are confirmed as such. 

In summary, we obtained 76 new redshift measurements of galaxies, 48 of which are 
projected onto the
Virgo cluster and 28 in the direction of the Coma--A1367 supercluster. 
With these new data, the redshift completeness in the VCC region remains 92\%
for $\rm B_T\leq16.0$ and 68\% for $\rm B_T\leq18.0$.
All membership estimates, as given in the VCC, are confirmed.
We remark that
a large fraction of the possible members classified as BCDs?, are found to be giant emission-line galaxies
well beyond the Virgo cluster.

The redshift completeness of CGCG galaxies in the direction of the  
Coma--A1367 supercluster is now 98\%.

\acknowledgements {We wish to thank the TACS of the Loiano, Cananea and OHP telescopes
for the generous amounts of time allocated to this project. G.G. wishes to thank his
students for their contribution during the observations and the
data reduction. 
This work could not be completed without the use of the NASA/IPAC Extragalactic Database (NED)
which is operated by the Jet Propulsion Laboratory, Caltech under contract with  NASA. 
L.C. has had support from CONACYT (M\'exico) research grant No. G-28586E.}


\begin{thebibliography}{}  

\bibitem[]{}
Binggeli B., Sandage A., \& Tammann G., 1985, AJ, 90, 1681
\bibitem[]{}
Binggeli B., Popescu C. \& Tammann G., 1993, AAS, 98, 275
\bibitem[]{}
Edmunds M., \& Pagel B., 1984, MNRAS, 211, 507
\bibitem[]{}
Haynes M., \& Giovanelli R.: 1986, ApJ, 306, 466 
\bibitem[]{}
Hoffman L., Helou G., Salpeter E., Glosson J., \& Sandage A., 1987, ApJS, 63, 247 
\bibitem[]{}
Hoffman G., Lewis B., Helou G., Salpeter E., Williams B. 1989, ApJS, 69, 65
\bibitem[]{}
Hoffman G., Lewis B., \& Salpeter E., 1995, ApJ, 441, 28
\bibitem[]{}
Gavazzi G., Boselli A., Scodeggio M., Pierini D. \& Belsole E., 1999a, MNRAS, 304, 595
\bibitem[]{}
Gavazzi G., Carrasco L., \& Galli R., 1999b, AAS, 136, 227 (G99)
\bibitem[]{} 
Grogin N., Geller M., \& Huchra J., 1998, ApJS, 119, 277 (GGH98)
\bibitem[]{}
Lemaitre G., Kohler D., Lacroix D., Meunier J., \& Vin A., 1990, A\&A, 228, 540
\bibitem[]{}
Magri C., 1994, ApJ, 108, 896
\bibitem[]{}
Sandage A., Binggeli B., \& Tammann G., 1985, AJ, 90, 395
\bibitem[]{}
Tonry J., and Davis M., 1979, AJ, 84, 1511 
\bibitem[1961-68]{Zwicky}
Zwicky F., Herzog E., Karpowicz M., Koval C.,  \& Wild P., 1961-1968, 
Catalogue of Galaxies and Clusters of Galaxies. (Pasadena: Caltech)(GCGC) 

\end{thebibliography}
\end{document}